\documentstyle[psfig,color]{mn}

\newcommand{\rmn}[1]{\mathrm {#1}}

\newcommand{\apj}[0] {ApJ}
\newcommand{\mnras}[0] {MNRAS}

\pagecolor[named]{White}

\pagerange{\pageref{firstpage}--\pageref{lastpage}}

\pubyear{2001}

\begin{document}
\label{firstpage}

\title[Convectively Unstable Outflows]
{Convective cores in galactic cooling flows}
\author[A. Kritsuk, T. Plewa \& E. M\"uller]{
Alexei Kritsuk$^{1,2}$, Tomasz Plewa$^{1,3}$ \& Ewald M\"uller$^1$\\
$^1$ Max-Planck-Institut f\"ur Astrophysik, Postfach
1317, D-85741 Garching, Germany\\
$^2$ Sobolev Astronomical Institute, University of St Petersburg, 
Stary Peterhof, 198904 St Petersburg, Russia\\
$^3$ Nicolaus Copernicus Astronomical Center, Bartycka 18, 00716
Warsaw, Poland}

\date{Submitted: August 1, 2000; Revised: March 21, 2001}

\maketitle

\begin{abstract}
We use hydrodynamic simulations with adaptive grid refinement to study
the dependence of hot gas flows in X-ray luminous giant elliptical
galaxies on the efficiency of heat supply to the gas.  We consider a
number of potential heating mechanisms including Type Ia supernovae
and sporadic nuclear activity of a central super-massive black hole.
As a starting point for this research we use an equilibrium
hydrostatic recycling model \cite{kritsuk96}.  We show that a compact
cooling inflow develops, if the heating is slightly insufficient to
counterbalance radiative cooling of the hot gas in the central few
kiloparsecs.  An excessive heating in the centre, instead, drives a
convectively unstable outflow.  We model the onset of the instability
and a quasi-steady convective regime in the core of the galaxy in
two-dimensions assuming axial symmetry.

Provided the power of net energy supply in the core is not too high,
the convection remains subsonic.  The convective pattern is dominated
by buoyancy driven large-scale mushroom-like structures.  Unlike in
the case of a cooling inflow, the X-ray surface brightness of an (on
average) isentropic convective core does not display a sharp maximum
at the centre. A hybrid model, which combines a subsonic peripheral
cooling inflow with an inner convective core, appears to be stable.
  We also discuss observational implications of these results.
\end{abstract}


\section{Introduction}
Extended thermal X-ray emission from the hot ($\sim10^7$~K)
interstellar medium (ISM) in giant elliptical galaxies is usually
interpreted in terms of a `cooling flow' scenario [see Loewenstein
\shortcite{1997gccf.conf..100L} for a recent review].  It implies that
radiative cooling of optically thin hot plasma drives a subsonic
inflow towards the centre of a potential well formed by the stellar
component and a massive dark halo.  Galactic cooling flows and cooling
flows in clusters of galaxies are essentially different owing to a
difference in the origin of the two media.  The intracluster medium is
mostly primordial, cools from its high virial temperature and is
accreted by the central galaxy, which provides a focus for the flow.
In contrast, the ISM in elliptical galaxies is constantly replenished
by mass loss from evolved stars.  This gas must be thermalized in the
galaxy's gravity field and heated to X-ray temperatures from which it
may cool down again.  Thus, in hydrodynamic terms, galactic cooling
flows are defined as systems where the mass flux is dominated by
source terms as opposed to `boundary terms' as in cluster cooling
flows \cite{1997gccf.conf..100L}.

Several potential heat sources have been considered to counterbalance
radiative cooling of the hot ISM in ellipticals.  The energy supplied
with normal stellar mass loss is limited by the value of the stellar
velocity dispersion and may only provide temperatures a factor of
$\sim2$ lower than the ISM temperatures \cite{davis.96}.  In contrast,
gravitational energy liberated in case of gas {\em inflow} would
provide too much heat to the central regions [Thomas
\shortcite{thomas86} and references therein].

Heating by supernovae (SN) Type Ia remains a controversial issue.  On
the one hand, the SN rate is not very well constrained.  Its value
varies from $0.55\:h_{75}^2$ SNu, suggested by van den Bergh \&
Tammann \shortcite{vandenbergh.91}, the uncertainty being a factor of
the order of 1.5, to $0.18\pm0.06\;h^2_{75}$~SNu derived by
Cappellaro, Evans \& Turatto
\shortcite{cappellaro..99}.\footnote{$h_{75}$ is the Hubble constant
in units of $75\,$km~s$^{-1}$Mpc$^{-1}$; 1~SNu = 1 supernova per
century per $10^{10}\,$ solar bolometric luminosities.}

On the other hand, the low iron content of the diffuse hot gas in
ellipticals estimated from high-quality broad band X-ray spectra
provided by ASCA for a single-temperature plasma model
\cite{1998IAUS..188...53L} suggests a very low efficiency of ISM
enrichment by Type Ia SNe.  This implies either that the SN rate is
lower than $\sim0.03\;h^2_{75}$~SNu \cite{1998IAUS..188...53L} or that
SN ejecta do not mix with the hot ISM
\cite{thomas86,1997ApJ...488..585F}.  However, the multi-parametric
analysis of X-ray spectra is a complex procedure based on iron L line
diagnostic tools, and requires accurate atomic physics data
\cite{1997ApJ...477..128A}.  The procedure is also model-dependent.
Using two-temperature multi-phase plasma models Buote
\shortcite{1999MNRAS.309..685B} obtained substantially better spectral
fits for the same data sets with iron abundances of $\sim1-2$ solar
and relative element abundances fixed at their solar values.  His
results are consistent with the Type Ia SN rate being up to a factor
of $\sim2$ lower than reported by Cappellaro et
al.~\shortcite{cappellaro..99}.  Clearly, better quality data are
required to determine the Type Ia SN heating rate more precisely.

All of the above mentioned energy sources may be described as being
continuously distributed within a galaxy and their local rates
depending on the stellar mass density and velocity dispersion, the gas
inflow velocity profile, and the shape of the gravitational potential.
There is a `global problem' \cite{1988cfcg.work..235T} of balancing
heating and cooling both in the centre and in the outer regions, since
the source terms depend in different ways on physical characteristics
which vary with radius.  However, empirical scaling laws for
ellipticals and the physics of radiative cooling and thermal
instability imply certain restrictions on the radial dependences.  
One possible solution to this problem is a hydrostatic hot gas
recycling model \cite{kritsuk96} for hot coronae of elliptical
galaxies. In the hydrostatic gas configuration all that remains of the
gas dynamical equations are the algebraic source terms which balance
to zero.  In this model two equilibrium conditions are simultaneously
satisfied: (1) the stellar mass loss rate exactly balances the rate
that mass cools locally from the flow (dropout) and (2) the rate that
thermal energy is radiated away is exactly balanced by the energy
input from stars and supernovae. The recycling model can be used as a
tool to distinguish between inflow and outflow regimes for a given set
of galaxy parameters (Kritsuk, B\"ohringer \& M\"uller
1998)\nocite{kritsuk..98}.

Periods of sporadic activity can drastically change the thermal state
of the ISM in the inner `cooling flow' region of a cluster or of an
elliptical on a time scale of $\sim10^8$~yr.  There are numerous
observations of an interaction between radio sources and the hot gas
both in central cluster galaxies and other ellipticals
\cite{harris...1999,1999ApJ...525..621K,2000ApJ...534L.135M}.  It is
this complex central region that is most sensitive to any imbalance in
the energy budget of the ISM.  Since the `thermal' time scale is
shorter there, the core of a few kpc in radius is the first to reach a
new equilibrium state corresponding to the current status of heating.
High spatial and spectral resolution X-ray observations of these
central regions are critical to distinguish among the various heat
sources and flow regimes in the core.

The response of a cooling flow to energy injection by the central
black hole that is fed by the cooling flow was simulated by Binney \&
Tabor (1995).  In their spherically symmetric numerical models the
black hole heats the central kiloparsec of the ISM as soon as it
begins to swallow gas from the ISM.  The resulting expansion of the
heated gas eliminates the cusped density profile and establishes a
core that is roughly 2~kpc in radius.  After the central heat source
has switched off, the core again cools catastrophically within
0.5~Gyr.  Among several limitations of the model (which include a
peculiar feedback mechanism) the authors particularly note the
restriction of spherical symmetry.  Both the jet and the ensuing
convective instabilities cannot be properly modelled due to this
symmetry restriction.

In this paper we analyze the hot gas flow regimes, which develop in
elliptical galaxies in response to a small imbalance in the energy
budget caused by a global instability of the equilibrium recycling
model \cite{kritsuk..98}.  We use multi-dimensional numerical
simulations to investigate compact cooling inflows and subsonic
convection -- two alternative saturated nonlinear regimes for a
slightly disturbed thermal equilibrium in the core.  Here, we restrict
ourselves to spherically symmetric (on the average) convective flows,
which have not been studied before.  The symmetry restriction will be
relaxed in a subsequent paper which more closely addresses the
astrophysical aspects of the problem.  The analysis presented here
provides a basis for a better understanding of more complex convective
flows and allows for a direct comparison with one-dimensional models.

A short description of our basic model is given in Section 2.  Section
3 describes the numerical method.  The results of numerical
experiments are presented in Section 4 together with the detailed
properties of the inner convective cores.  In Section 5 we summarize
the results and discuss effects to be expected for the dynamical state
of the hot gas in ellipticals from a larger imbalance between cooling
and heating.

\section{The model}
Our numerical experiments are based on a physical model, which has
been described in Kritsuk et al. \shortcite{kritsuk..98}.  We refer
the reader to that paper and references therein for a more detailed
description than given below.

We assume that the local mass budget of the hot ISM inside the galaxy
is determined exclusively by mass supply from the old stellar
population of the galaxy [in the form of stellar winds, planetary
nebulae, supernova explosions, etc.; a detailed discussion can be
found in Mathews \shortcite{mathews90} and Mathews \& Brighenti
\shortcite{mathews.99}], and by gas condensation to a warm phase due
to thermal instabilities.  Our motivation for including mass dropout
is based on the fact that the assumed mass supply mechanisms
inevitably cause nonlinear small scale perturbations in the hot phase,
which can trigger thermal instabilities.  The rate of mass dropout is
assumed to be a function of local conditions in the hot ISM, and given
by the expression $\dot\rho_{\rmn{ti}} = b\chi(n)\rho$, where $\rho$
is the density of the hot phase.  The spectrum of the initial
small-scale perturbations in the gas and, in particular, its
modification by heat conduction are described by a single parameter $b
\equiv const\in [0,\; 1]$.  The Heaviside function $\chi$ and the
linear instability growth rate $n$ are defined as
\begin{equation}
\chi(n)=\cases{n, &if $n\ge 0$;\cr0, &otherwise;} 
\end{equation}
\begin{equation}
n\equiv {\partial \over \partial t} 
\left(\ln{\delta \rho \over \rho}\right)={1\over
c_p}\left({2\rho\Lambda\over T}-\rho{\rmn{d}\Lambda\over \rmn{d} T}\right) -
{\alpha\rho_*\over \rho}. \label{GrowthRate}
\end{equation}
Here $T$ is the hot gas temperature, $\Lambda(T)$ is the
nonequilibrium zero field radiative cooling function for solar
composition and complete ionization of the plasma \cite{SD93}, and
$c_p$ is the specific heat at constant pressure.  The first term in
the {\em rhs} of equation (\ref{GrowthRate}) coincides with Field's
instability criterion, and the second one describes stabilization due
to stellar mass loss \cite{kritsuk92}, $\alpha$ is the specific
stellar mass loss rate, and $\rho_*$ is the local stellar mass
density.  Here we assume $b\approx 0.5$ \cite{kritsuk96}.  Since we
are interested in relatively short evolutionary time scales ($\le
1$~Gyr), we assume $\alpha$ to be a constant.

The local energy budget of the hot ISM is determined by radiative
losses and energy supply from stars (via SN Type Ia explosions and via
thermalization of the kinetic energy of orbital motion for the gas
shed by stars).  Radiative cooling comprises two energy sinks due to
(i) direct radiative cooling of the hot phase and (ii) due to
radiative losses by the condensing material.

Finally, the hydrodynamic description for the hot phase is based upon
the assumption that the sources are homogeneously distributed.  Note
that this assumption is fulfilled as long as the numerical resolution
is not excessively high.  The set of conservation laws for mass,
momentum, and energy of the hot phase can then be written as follows:
\begin{equation}
\frac{\partial \rho}{\partial t} + \bmath{\nabla\cdot}(\rho \bmath{v})=\alpha
\rho_* - \dot\rho_{\rmn{ti}}, \label{Mass}
\end{equation}
\begin{equation}
\frac{\partial (\rho \bmath{ v})}{\partial t} + 
\bmath{v\nabla\cdot}(\rho\bmath{v}) +
\rho\bmath{v\cdot\nabla v} +
\nabla p=\rho\nabla\phi - \dot\rho_{\rmn{ti}}
\bmath{v}, \label{CompleteSystem}
\end{equation}
\begin{eqnarray}
\lefteqn{\frac{\partial E}{\partial t}+\bmath{\nabla\cdot}[\bmath{v}(E+p)] = } 
\nonumber \\ 
& & \alpha\rho_* e_* -\dot\rho_{\rmn{ti}}(E+p)/\rho-\rho^2\Lambda +\rho
\bmath{v\cdot\nabla}\phi. \label{Energy}
\end{eqnarray}
Here $\bmath{ v}=(u,v,w)$ is the velocity vector, $p=(\gamma-1)\rho e$
is the pressure, $\gamma$ is the adiabatic index, $e$ is the specific
internal energy, and $e_* = c_{\rmn{v}} T_0$ is the characteristic
specific energy of the heat source of temperature $T_0$.  The energy
density $E=\rho(e+\bmath{v}^2/2)$.

We define a spherically symmetric gravitational potential $\phi$ to be
that of a two-component isothermal sphere, which includes stars and a
massive dark halo \cite{kritsuk97}.  The galaxy model is characterized
by four parameters: the stellar velocity dispersion $\sigma_*$,
the characteristic radius of the stellar mass distribution
\begin{equation}
r_0=\frac{\sigma_*}{\sqrt{4\pi G \rho_{*,0}}}, \label{king_rad}
\end{equation}
the ratio of the stellar and dark matter (DM) velocity dispersions
\begin{equation}
\beta\equiv{\sigma_*^2\over\sigma_{\rmn{DM}}^2}<1,
\end{equation}
and the ratio of the dark matter mass density and the stellar mass
density at the centre of the galaxy
\begin{equation}
\delta=\rho_{\rmn{DM},0}/\rho_{*,0}.
\end{equation}
For the prototype galaxy NGC~4472, we set $\sigma_*=304$~km s$^{-1}$,
$r_0=0.16$~kpc, $\beta=0.5$, $\delta=10^{-1.5}$, which allows us to
reproduce its surface brightness profiles both in the optical and in
X-rays with a reasonable accuracy.  In this case sinks and sources of
mass and energy for the hot phase are locally compensated everywhere
in the galaxy if the parameters $\alpha$, $b$, and $e_*$ are constant.
Hydrostatic equilibrium requires an adjustment of the temperature of
the hot gas $T_{\rmn{eq}}(b, T_0)$ and of the stellar velocity
dispersion $\sigma_*$. This is achieved by additional restrictions
applied to the values of $b$ and $T_0$ coupled by
\begin{equation}
{{\cal R}\over \mu} T_{\rmn{eq}}(b, T_0) = 2\sigma_*^2,
\end{equation}
where ${\cal R}$ is the gas constant and the mean molecular weight
$\mu=0.62$ [see \cite{kritsuk..98} for more detail].  For a specific
stellar mass loss rate $\alpha=5\times10^{-20}$~s$^{-1}$
\cite{faber.76b,sarazin.87}, hydrostatic equilibrium is achieved with
a reasonable mass deposition efficiency of $b=0.48$ and a
characteristic temperature of the heat source of $T_0=5\times10^7$~K.
This temperature is lower than the $7\times10^7$~K corresponding to a
SN Type Ia rate of $0.55h^2_{75}$~SNu \cite{vandenbergh.91}, but it is
higher than the $3\times10^7$~K obtained for a supernova rate of
$0.18\;h^2_{75}$~SNu \cite{cappellaro..99}.

Note, that we do not claim NGC~4472 to be a {\em perfect} galaxy with
an isothermal hot ISM, an isothermal stellar system and an isothermal
dark halo.  Rather we construct a simple model whose physical
properties are based on observations of NGC~4472.  The use of a
restricted set of control parameters allows us to investigate in which
way(s) our model differs from the real object.

\section{Numerical method}
The evolutionary equations are solved by means of the AMRA code
\cite{PM00}, which is based on the direct Eulerian version of the PPM
method of Colella \& Woodward \shortcite{CW84} on an adaptively
refined hierarchical system of completely nested grids
\cite{BC89}. For the present 1D and 2D axisymmetric simulations we use
spherical coordinates. The computational domain extends from
$r_{\rmn{in}} = 100$\,pc ($r_{\rmn{in}} = 0$\,pc in 1-D simulations)
to $r_{\rmn{out}} = 400$\,kpc in radius and covers (in 2D simulations)
the wedge $10^{\circ} \le \theta \le 170^{\circ}$.  Periodic boundary
conditions are imposed in the angular direction.\footnote{Unlike
reflecting, periodic boundary conditions allow for a free gas flow
across the sides of the computational volume, which is closer to the
3-D case. Switching from periodic to reflecting conditions in the
steady convective regime does only modify the extreme values of the
velocities. The main flow remains basically unchanged.}  In the radial
direction hydrostatic equilibrium is assumed at $r_{\rmn{in}}$ and at
$r_{\rmn{out}}$, i.e.  the radial (and theta) velocity is set to zero
and the boundary pressure is defined in such a way that the pressure
gradient at the boundary is equal to the hydrostatic one.  We allow
for up to 4 levels of refinement with density and pressure jumps of 10
per cent used as refinement criteria. With refinement ratios of 4 and
2 in radius and angle, respectively, the resolution of our 2-D models
is equivalent to that of a uniform grid run with $32768 \times 160$
zones ($\Delta r \simeq 12$\,pc and $\Delta \theta = 1^{\circ}$).  The
source terms are calculated explicitly \cite{P93} using the operator
split approach.

\section{Results}
\begin{figure*}
\vspace{-4.7cm}
\psfig{figure=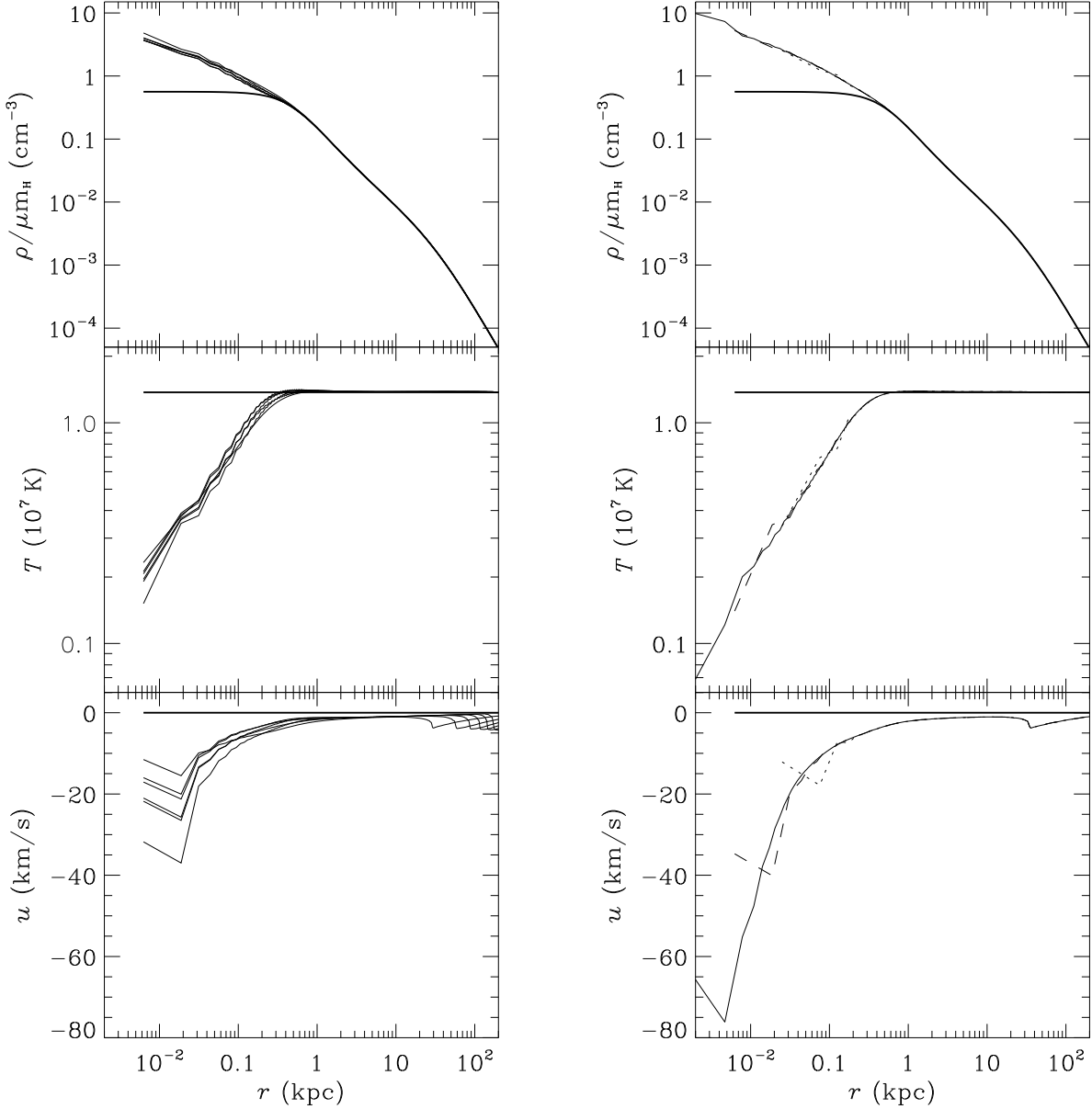,clip=}
\vspace{-4.cm}
\caption{Quasi-steady state inflow (initial density perturbation
$\varepsilon=0.1$).  
The left panel shows the density, temperature and velocity distributions
at $t=0$, 50, 100, 150, 200, 250, and 300~Myr for an inflow solution 
obtained with intermediate spatial resolution of 12~pc.
The right panel shows hydrodynamic variables
at $t=60$~Myr.
The solutions obtained with a spatial resolution of 3, 12, and 48~pc are 
shown by solid, dashed, and dotted lines, respectively. 
The heavy lines refer to the initial conditions.
}
\label{Inflow}
\end{figure*}

\subsection{Spherically symmetric solutions}
In the presence of heat conduction the equilibrium recycling model for
the hot gas in giant ellipticals is meta-stable \cite{kritsuk..98}.
It means that, while arbitrary infinitesimal perturbations do not
grow, perturbations of a finite amplitude can switch the system to a
different equilibrium state or to a nonequilibrium catastrophic
evolutionary regime.

Since we are mostly interested in studying thermal effects and since
the characteristic thermal time scale is usually longer than the
dynamical time, it is quite natural to examine the response of the
system to a class of perturbations, which do not disturb the
hydrostatic equilibrium.  In general, the system can be perturbed in a
number of different ways.  However, the approximation of spherical
symmetry, which allows us to reduce the problem to 1-D, is rather
restrictive.

The simplest perturbation of this class is a global density
perturbation $\delta\rho = \varepsilon \rho_{\rmn{eq}}(r)$ which, at
first sight, may seem to be unphysical because of its global nature.
However, due to stratification of the hot gas, the thermal time scale
depends strongly on radius, i.e. for a time scale of a few $10^8$~yr
the perturbation only affects the inner region because the rest of the
hot gas has not enough time to react.  Hence, for a sufficiently short
time interval the global perturbation is equivalent to a more local
one.

\subsubsection{Cooling inflows}
We have computed a series of spherically symmetric solutions with
initial global density perturbation amplitudes $\varepsilon=0.1$ and
0.2, and compared the results obtained with different spatial
resolution in the central cooling region.  Left and right panels of
Fig.~\ref{Inflow} illustrate our results for $\varepsilon=0.1$.  The
left panel shows distributions of density temperature and velocity at
$t=0$, 50, 100, 150, 200, 250, and 300~Myr for our intermediate
resolution run with a radial spacing of 12 pc. The heavy lines
indicate the initial conditions.  The quasi-steady regime is already
established at $t\approx50$~Myr after having passed through the linear
growth regime which covers most of the earlier epochs. Then, for a
very short time, a transient dynamic inflow occurs. Afterwards the
flow remains quasi-steady.  The gas temperature in the cooling core
drops below the equilibrium value $T_{\rmn{eq}}$ by a factor of
$\sim10$.  The density distribution shows a central cusp $\rho\propto
r^{\nu}$ with a power index $\nu = -0.6$.  Hence, this model predicts
a compact central peak in the X-ray surface brightness. This peak can
be observed if the instrument has sufficient angular resolution, and
if finite optical depth effects in the inflowing dense core will not
render the approximation of an optically thin plasma invalid.  Note,
that heat conduction and/or magnetic pressure support can modify the
stability of the initial equilibrium model and even suppress the
formation of the cooling core \cite{kritsuk..98}.

The right panel shows the flow variables at $t=60$~Myr soon after a
compact cooling inflow has been established in the core region of
$\sim1$~kpc in radius.  The effects of the finite grid resolution are
most evident in the velocity plot.  Changes in the grid resolution
effectively modify the position of the inner boundary thus determining
the lowest velocity and temperature, and the highest density values
attained on the grid.

\begin{figure}
\vspace{0.cm}
\psfig{figure=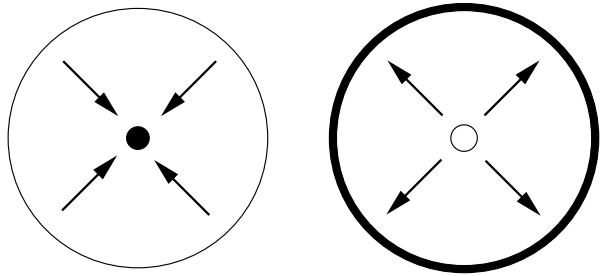,clip=}
\vspace{-8.cm}
\caption{Sketch of the two alternative solutions for the meta-stable
hydrostatic equilibrium in spherical symmetry: a cooling {\em inflow}
(left) and a subsonic {\em outflow} (right).  The regions of intense
mass deposition are shown in solid black in the vicinity of the centre
in case of inflow and a condensation front in case of outflow.}
\label{Sketch}
\end{figure}

At the coarsest grid spacing of 48~pc the compact inflow region ($100
\la r \la 500$~pc) is covered by less than 10 grid zones and hence
remains unresolved.  Steep gradients of dependent variables introduce
interpolation errors, and errors due to explicit calculation of the
source terms.  These errors caused by the finite difference scheme are
amplified by the physical thermal instability and, as a result, for
$\varepsilon = 0.1$ the numerical solution displays high frequency
noise in the temperature distribution for $t \ga 65\,$Myr.  Using the
same spatial resolution but doubling the size of the initial
perturbation ($\varepsilon = 0.2$) makes the inflow even more dynamic
and the noise amplitude becomes so large that the numerical solution
eventually switches from the inflow to the outflow regime (see
Fig. \ref{Sketch}).

Grid resolution determines the amount of numerical dissipation of PPM.
Although PPM does not utilize artificial viscosity explicitely, its
interpolation scheme introduces some amount of dissipation, which
depends on grid resolution.  When increasing the resolution by a
factor of 16 to $\Delta r=3$~pc the decreased numerical dissipation is
found to be too small to prevent the development of small scale
condensations due to the physical thermal instability inherent to
flows with radiative cooling [see, e.g., Hunter \shortcite{hunter70},
Binney \& Tabor \shortcite{binney.95}].  As a result, a quasi-steady
inflow solution is obtained (solid lines in the right panel of
Fig.~\ref{Inflow}), which becomes unstable soon after $t=200$~Myr
causing the breakup of the central density enhancement into multiple
spherical condensation fronts.

At an intermediate spatial resolution of 12~pc (left panels and dashed
lines in the right panels of Fig.~\ref{Inflow}) the cooling inflow
region is well resolved by $\sim 32$ grid zones. The numerical
solution remains stable for at least 500~Myr beyond the initial
relaxation to a steady state inflow at $\sim 55$~Myr.  Note that at
higher spatial resolution (solid lines; right panels) the profiles are
smoother than at lower grid resolution (left panels). The 'ripples'
present at lower resolution can be eliminated, for instance, by
including a small phenomenological diffusion term in the equations, as
in Binney \& Tabor \shortcite{binney.95}, or by taking into account
thermal conductivity [Kritsuk et al.  \shortcite{kritsuk..98}].

The numerical experiments discussed below are all performed with a
minimum radial grid spacing of 12~pc. At a much coarser grid important
hydrodynamic scales remain unresolved, whereas a much finer grid is no
longer consistent with our basic assumption that the sinks and sources
of mass and energy are distributed continuously.

Preliminary results obtained in two dimensions assuming axial symmetry
indicate that cooling inflows are stable against random small velocity
perturbations.  The stability is maintained due to advection of linear
perturbations towards the inner boundary by the accelerating gas
inflow.

\subsubsection{Subsonic outflows}
\begin{figure*}
\vspace{-4.5cm}
\psfig{figure=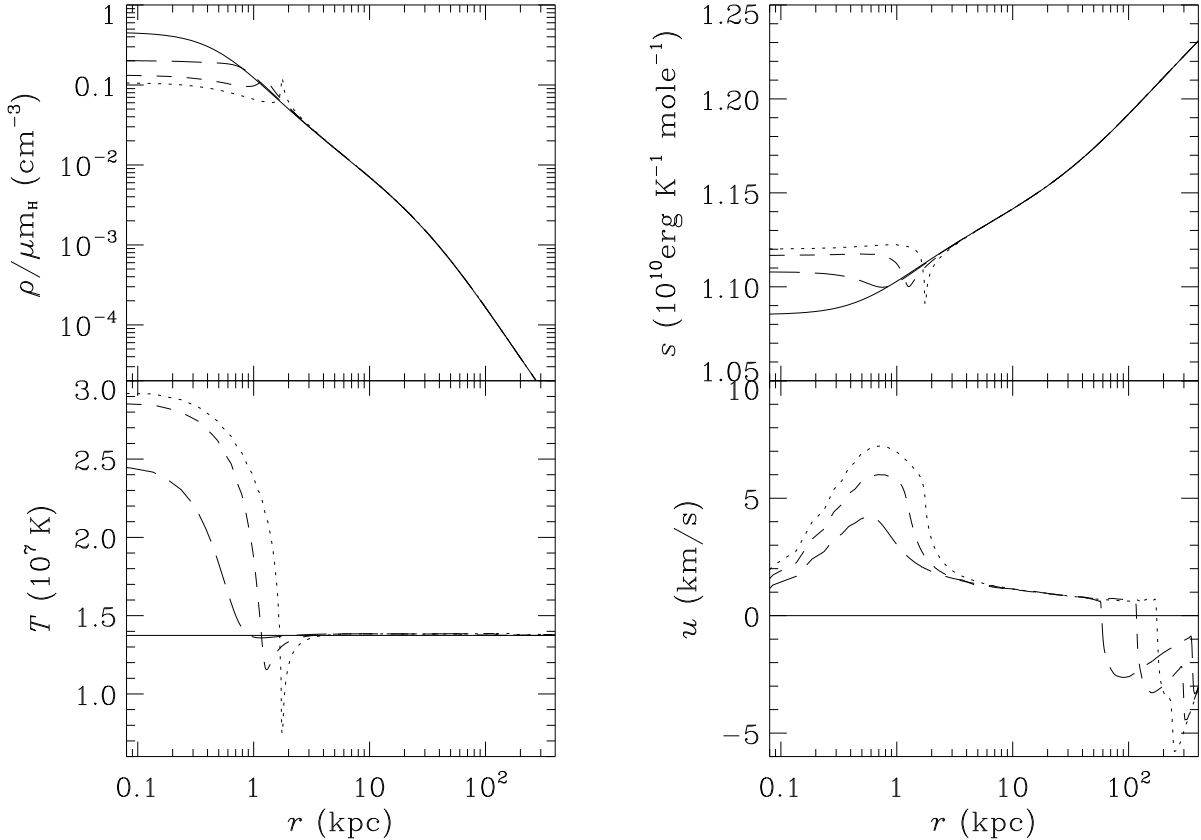,clip=}
\vspace{-9cm}
\caption{A subsonic outflow solution (initial density perturbation
$\varepsilon=-0.1$).  The four panels show the density, temperature,
entropy and velocity distributions at $t=0$ (solid), 100
(long-dashed), 200 (dashed), and 300~Myr (dotted), respectively.}
\label{Outflow}
\end{figure*}
As an alternative to quasi-steady cooling inflows we have studied a
class of subsonic outflow solutions, which can be obtained through
negative global perturbations added to the equilibrium density.  In
such a case heating and mass supply from evolved stars in the galaxy
overtake cooling and mass dropout, and instead of a compact cooling
core a hot bubble develops in the centre (Fig.~\ref{Sketch}).

Fig.~\ref{Outflow} shows an example of a subsonic outflow solution at
$t=0$, 100, 200, and 300~Myr for $\varepsilon=-0.1$.  In addition to
the density, temperature, and velocity distributions we also show the
entropy, which is defined here as
\begin{equation}
s = c_{\rmn{v}} \ln\left(\frac{T}{\rho^{\gamma-1}}\right) + const.
\end{equation}
Numerical values for the entropy $s$ are given assuming $const=0$.

As the gas in the hot bubble expands it compresses the ambient medium
and thus creates a leading spherical quasi-isobaric condensation wave.
The wave front is clearly seen as a peak in density and as a sharp
minimum in the temperature profiles.  The gas temperature in the
bubble is a factor of $\sim2$ higher than the equilibrium temperature
$T_{\rmn{eq}}$. It eventaully saturates at a value of about $3\times
10^7$~K which is fixed by the heating rate at the centre.  Typical
expansion velocities in the bubble are about 10 km s$^{-1}$. The
maximum velocity continuously grows as the bubble expands while less
and less dense outer layers of the hydrostatic corona start to feel
the energy imbalance due to the initial global density perturbation.
The bubble will expand beyond 300~Myr (the state shown in
Fig.~\ref{Outflow}) because of the special form of the initial
perturbation.  Thus, the solution will not tend towards a steady
galactic wind.  Since on a timescale of a few Gyrs the heating rate
and stellar mass loss vary, such steady wind solutions are
unrealistic, anyway.  Negative velocities observed at $r>50$~kpc are
related to a sound wave, which is triggered at the beginning of the
simulation by small deviations of the initial model from an exact
hydrostatic state.

The outflow solution is Rayleigh-Taylor unstable since the initially
stable entropy stratification is modified by the expanding bubble.  In
particular, one may expect that the unstable local entropy inversion
associated with the forming condensation front will be smeared out by
convective motion, if the restriction of spherical symmetry of the
flow is relaxed, and, thus, such a sharp front will never form in
reality.

\subsection{Unstable axisymmetric outflows}
In order to initialize the flow in the axisymmetric case we impose
besides a $\varepsilon=-0.1$ global density perturbation also a small 
random velocity perturbation with an amplitude of $10^{-3}$ of the adiabatic
sound speed. This additional axisymmetric perturbation helps us to
initiate two-dimensional flow. 
In the central region, $0.1\,{\rm kpc} \le r \le4\,$kpc,
the grid resolution is $320 \times 160$ zones.  The evolution has been
followed for 300~Myr.

\subsubsection{Flow structure and evolution\label{flowEvol}}
\begin{figure*}
\vspace{-9.5cm}
\psfig{figure=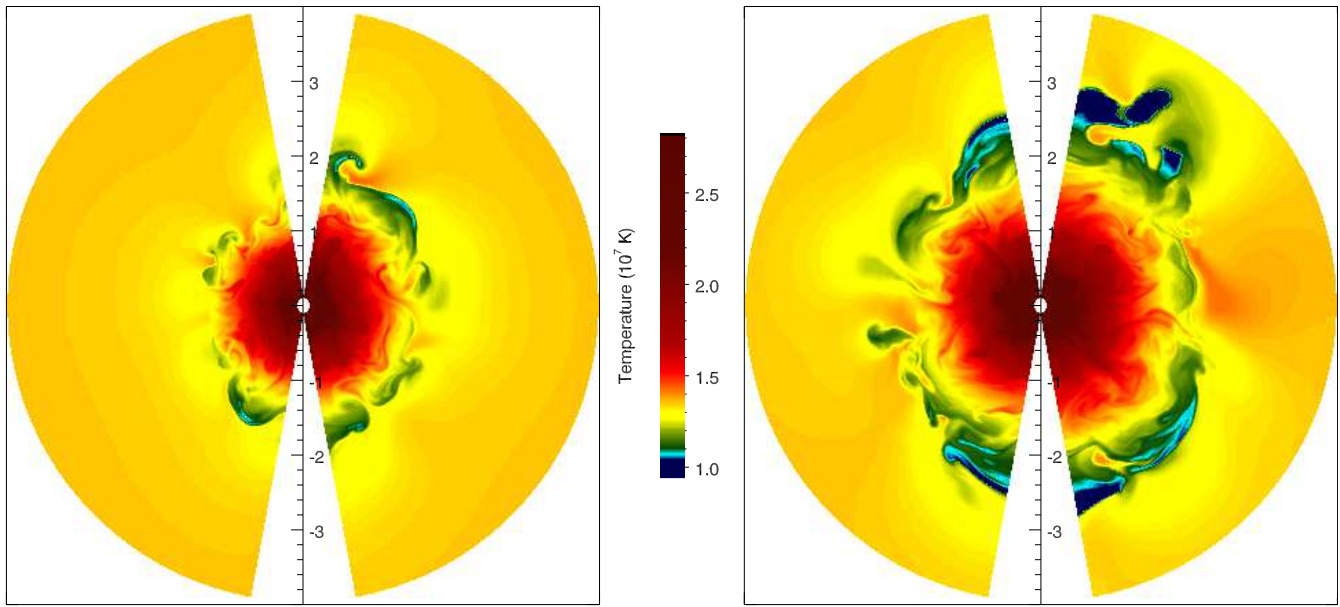,clip=}
\vspace{0cm}
\caption{Four snapshots of the temperature field in the innermost
4~kpc region (from left to right) taken at $t=150$, 200, 250, and
300~Myr, respectively.  The vertical scale along the symmetry axis
shows the radial distance in kiloparsecs.  Note that the left sectors
in both panels ($t=150$ and 250~Myr) are mirrored with respect to the
right ones.  The hot bubble at the centre can be seen in grades of red
and the leading condensation front smeared by convection in blue.}
\label{Temperature}
\end{figure*}
\begin{figure*}
\vspace{-9.5cm}
\psfig{figure=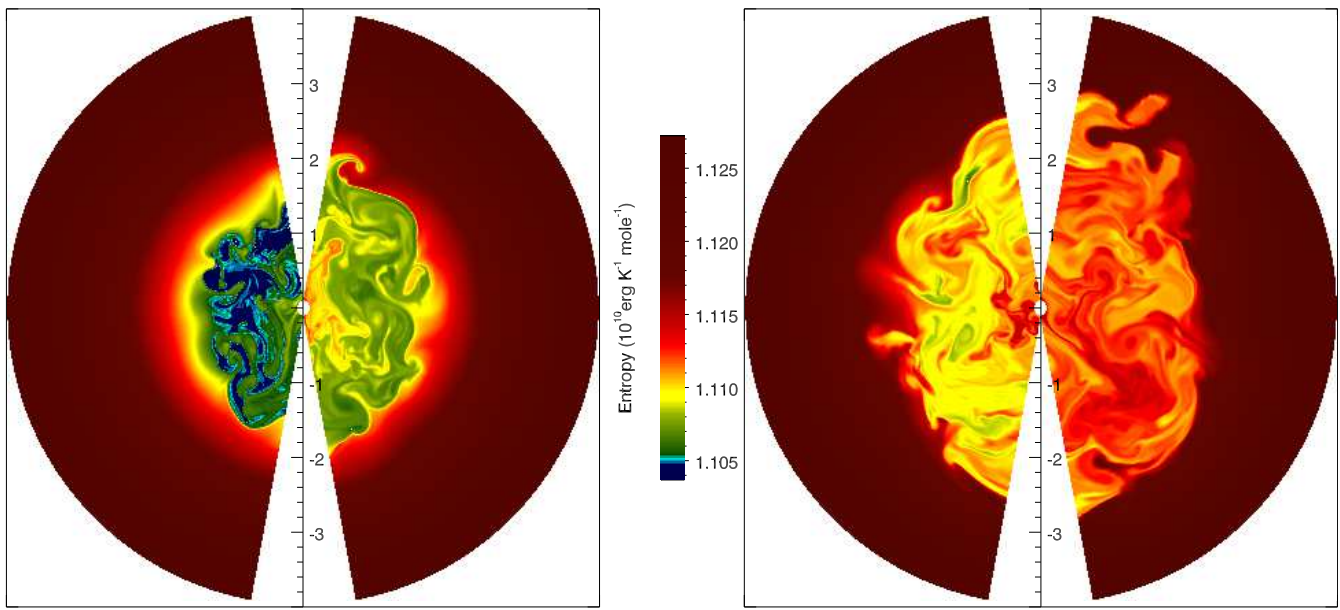,clip=}
\vspace{0.cm}
\caption{Same as Fig. 4 but for the entropy field.  Gas motions
triggered by the convective instability tend to level the inner part
of the entropy distribution.  A characteristic flow pattern in the
core consists of ascending and descending mushroom-like structures.}
\label{Entropy}
\end{figure*}
\begin{figure*}
\vspace{-9.5cm}
\psfig{figure=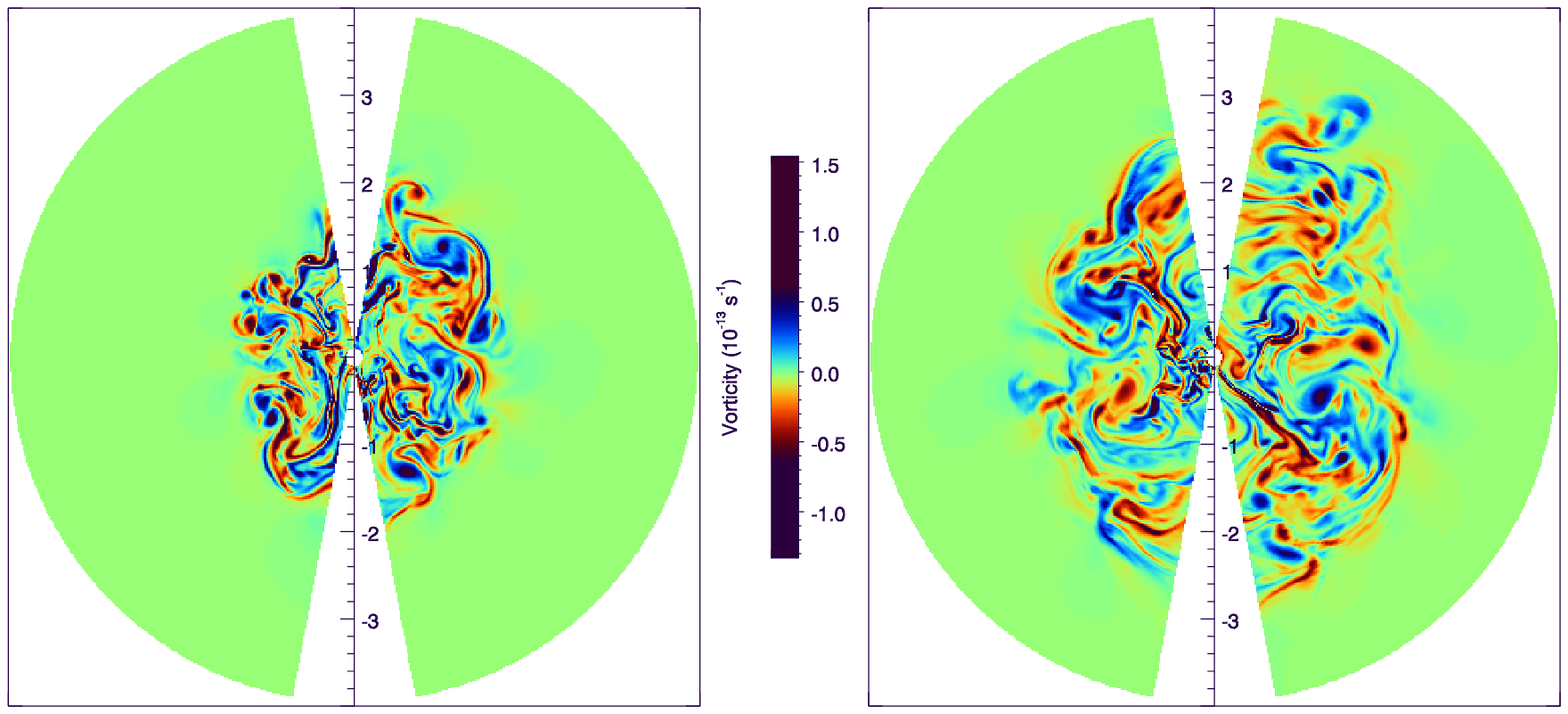,clip=}
\vspace{0.cm}
\caption{Same as Fig. 4 but for the vorticity field.  Vorticity as a
differential diagnostic of the velocity field highlights smaller scale
structures in the flow pattern.  Counter-rotating eddies can be seen
in blue and in red.}
\label{Vorticity}
\end{figure*}
 
Convection starts to develop after $t\approx60$~Myr from the beginning
of the simulation in the form of small-scale vertical finger-like
structures.  The fingers originate the central region of unstable
entropy inversion.  High entropy material flows upwards and low
entropy material -- downwards, restoring a marginally stable
stratification with an adiabatic temperature gradient
\begin{equation}
\frac{\rmn{d}T}{\rmn{d}r} = \left(1-\frac{1}{\gamma}\right)\frac{T}{P}
\frac{\rmn{d}P}{\rmn{d}r}.
\end{equation}
As the fingers flowing in opposite directions accelerate,
Kelvin-Helmholtz roll-up transforms them into mushroom-like structures
typical for the developing convective pattern \cite[Section
6]{inogamov99}.

By $t=100$~Myr a convective core with a radius of $\sim1$~kpc forms
and large scale structures become dominant.  At that time the radial
velocities of the convective motion reach $\pm 100$~km~s$^{-1}$, and
convection approaches a quasi-stationary regime.  Still later, when
the outer gas layers start to feel the imbalance of mass and energy
gains and losses due to the global initial density perturbation, these
layers become involved into the convective motion, too.  As the
temperature in the hot bubble grows and the convective core expands,
large scale convective structures with higher flow velocities appear.

At $t=300$~Myr the linear size of the largest mushroom is about 3~kpc
and typical flow velocities are $\sim 150$~km~s$^{-1}$.  A little bit
earlier, at $t=280$~Myr, the maximum Mach number of 0.54 is recorded
for a stream of upflowing hot gas.  Thus, the convection driven by
initially small\footnote{Note, that a negative perturbation of 
10 per cent in the gas
density implies an about 20 per cent lower luminosity than in
equilibrium, i.e. 20 per cent `extra' heating. This corresponds to a
net continuous supply of heat with a power of $\sim
10^{40}$~ergs~s$^{-1}$.  A conservative estimate of $\sim
10^{43}$~ergs~s$^{-1}$ for the typical power supplied by AGNs in the
form of mechanical energy of jets is much larger than this value (see,
e.g., Owen, Eilek \& Kassim \shortcite{owen..00}).  The typical power
required to sustain radio core sources in elliptical galaxies is about
$\sim 10^{39}$~ergs~s$^{-1}$.  Our simulations show that a smaller
($\varepsilon = -0.01$) initial density perturbation develops a
convective core with a radius of $\sim1.3$~kpc by 300~Myr.}
($\varepsilon = -0.1$) deviations from local thermodynamic equilibrium
remains subsonic.  In a realistic 3-D situation one could expect even
lower convective velocities due to the extra dimension available for a
redistribution of the kinetic energy.

Snapshots of the temperature, entropy and vorticity distributions for
the inner 4~kpc are shown in Figs \ref{Temperature}--\ref{Vorticity}.
The temperature maps clearly exhibit the expansion of the hot bubble
and the distorted analogue of the spherically symmetric leading
condensation front, which is formed by the uppermost parts of the
mushroom caps.  The mushrooms are best seen in the entropy maps owing
to the formation of an entropy `plateau' in the core.  The mean value
of the entropy in the core steadily grows. Consequently, the
convective pattern changes from blue via green and yellow to red.  The
vorticity maps highlight small scale structures in the velocity field
and help in tracing the mushroom clouds as counter-rotating eddies
forming the caps (Fig. \ref{Vorticity}).  The convective core as a
whole has a well defined sharp outer boundary defined by the largest
mushroom caps.  The global maximum of the hot gas density is typically
slightly off centre.

For further details on the flow evolution we refer to a set of
computer animations available at {\tt
http://www.mpa-garching.mpg.de/Hydro}.

\subsubsection{Statistical properties: angular averages}
\begin{figure*}
\vspace{-4.7cm}
\psfig{figure=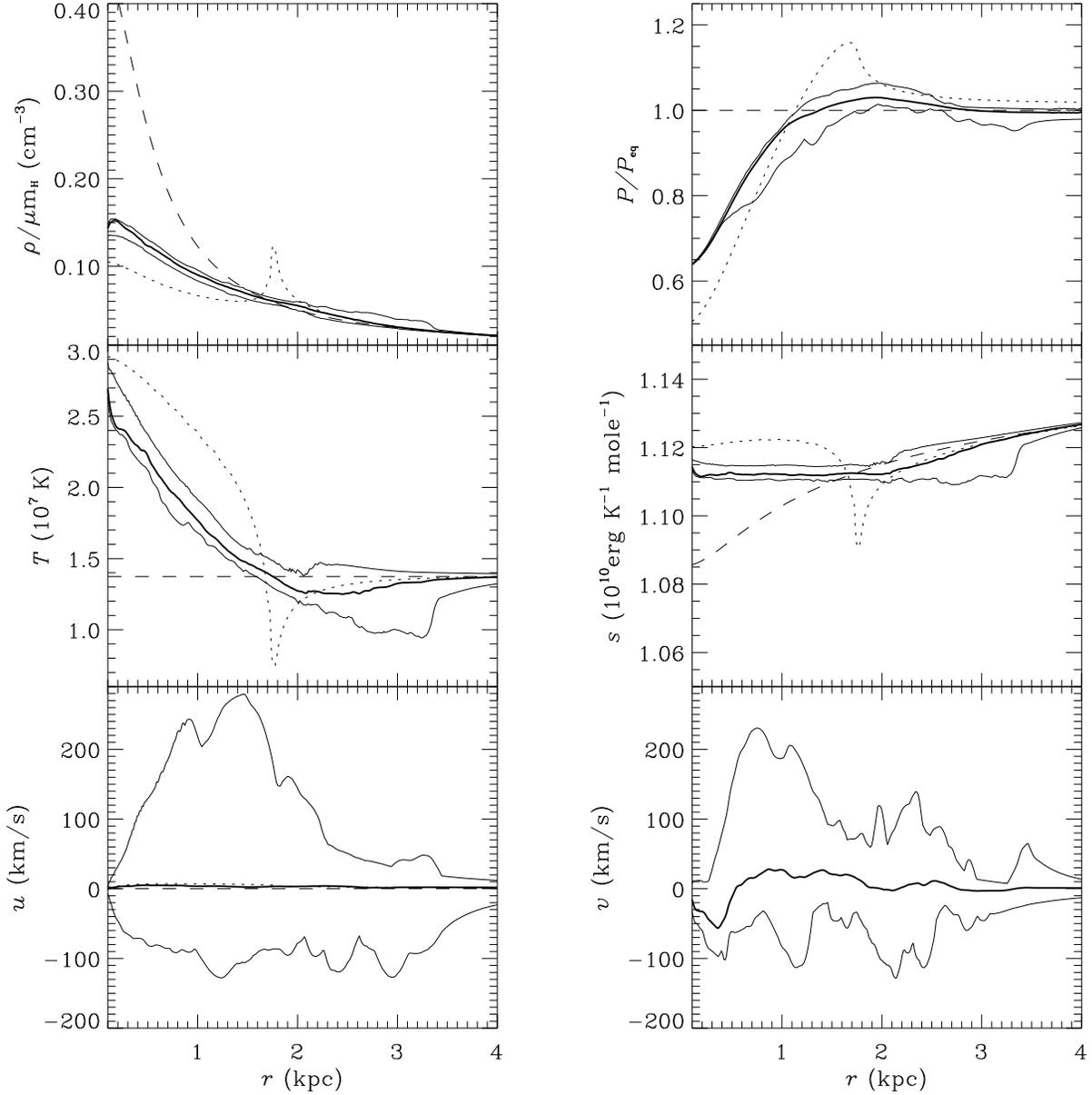,clip=}
\vspace{-4.cm}
\caption{Various hydrodynamic variables are shown as a function of
radius for $r\le4$~kpc at 300~Myr.  Heavy solid lines show angular
averaged values, solid lines show $max$ and $min$ envelopes (i.e., the
maximum/minimum angular value at the respective radius), dashed lines
give the initial conditions, and dotted lines show the spherically
symmetric solution at 300~Myr.  The pressure is normalized to the
initial hydrostatic value.  The gas temperature is the emission
measure weighted angular mean, while the velocity components (vertical
$u$ and horizontal $v$) and the entropy are mass weighted.}
\label{MeanFlow}
\end{figure*}
Analysis of the angular averages of the hydrodynamic variables allows
one to compare the properties of the new equilibrium\footnote{ We refer 
to the term `equilibrium' here in a statistical
sense, implying that the mean convective flow is quasi-steady on time
scales of a few 10~Myr. According to Fig.~\ref{Integral}, the flow
evolves slowly on a time scale of hundreds of Myrs which is longer
than the typical life span of a convective element (See Sections
\ref{fourier}, \ref{integral}).  }
established in the convective core with those of the hydrostatic
recycling model, and those of the outflow solution.

Fig. \ref{MeanFlow} shows angular averages at 300~Myr for the central
4~kpc region.  Convection has smeared out the condensation front,
which forms in the spherically symmetric case.  As a result, the
averaged convective solution falls in between the equilibrium model
and the 1-D outflow.

\begin{figure}
\vspace{-4.7cm}
\psfig{figure=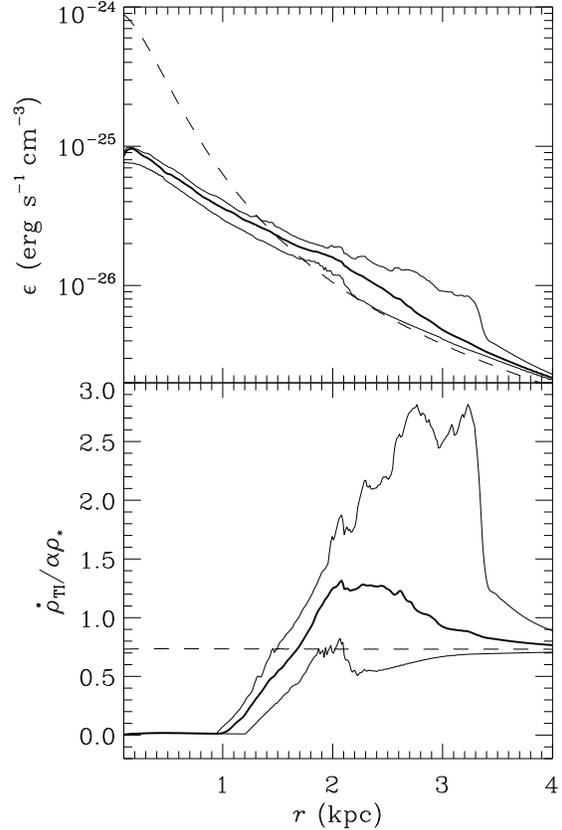,clip=}
\vspace{-9cm}
\caption{Mass and energy supply rates for the region $r\le4$~kpc at
300~Myr.  The upper panel shows the volume emissivity. The lower panel
displays the relative rates of mass supply to and dropout from the hot
phase.  Heavy solid lines show the angular averaged mean values, solid
lines are the $max$ and $min$ envelopes (i.e., the maximum/minimum
angular value at the respective radius), and dashed lines give the
initial conditions.}
\label{MeanSrc}
\end{figure}

The gas density near the very centre is by a factor of $\sim3$ smaller
than in the equilibrium, but still larger by a factor of $\sim1.5$
than in the 1-D case.  Unlike in a compact inflow, there will be no
sharp peak in the X-ray brightness in the central 1~kpc in this case
(see also the emissivity distribution in the upper panel of
Fig. \ref{MeanSrc}).  An observation with sub-arcsec resolution of a
nearby elliptical galaxy, like NGC~4472, would be a critical
experiment to distinguish between the compact inflow and convective
outflow realizations in the core.  The relatively small density
fluctuations about the mean indicate that compressibility is not very
important, which is consistent with a subsonic convection regime.

The average temperature grows by a factor of $\sim2$ in the central
1~kpc and has a shallow minimum at the position of the condensation
front.  Local relative temperature variations in the convective flow
are about $1.5$, mainly at the interface between the convective core
and the ambient hydrostatic gas.  The average pressure profile shows
clear deviations from the hydrostatic stratification.  Local pressure
fluctuations are $\le 5$~per cent, and hence imply a quasi-isobaric
regime of convection. The angular averaged entropy distribution is
almost flat within a 2~kpc core and rises further outwards.  There is
no entropy inversion as in the 1-D case, i.e. the system has settled
to a new equilibrium state via nonlinear saturation of the convective
instability.

Both vertical and horizontal gas velocities are quite high compared to
typical cooling flow values of about $-20$~km~s$^{-1}$.  However, the
mean radial velocity at $t=300$~Myr lies in the range $-1.0 \le u \le
5.2\,$km\,s$^{-1}$, which is similar to the 1-D case, where
$u_{\rmn{max}} (\le 4\, {\rmn kpc}) = 7.2\,$km\,s$^{-1}$.  Thus, there
is no considerable mass flux in the radial direction.
 
In order to assess the efficiency of radiative cooling in the
turbulent core, we define the volume emissivity $\epsilon$ as
\begin{equation}
\epsilon = \dot\rho_{\rmn{ti}}e + \rho^2\Lambda\,.
\label{Emiss}
\end{equation}
The two terms on the {\em rhs} in the above equation describe
radiative losses from the cooling condensations and direct radiative
cooling of the hot gas phase, respectively.  The upper panel in
Fig. \ref{MeanSrc} shows the distribution of the emissivity with
radius for the inner 4~kpc.  The peak central emissivity in the
convective core is an order of magnitude lower than in hydrostatic
equilibrium.  For $r \ge 1\,$kpc the emissivity distribution roughly
follows the one of the recycling model.  Local variations of the
emissivity by a factor of $\sim 3$ are detected in the condensation
front region between 2 and 3.5~kpc.

The lower panel in Fig.\,\ref{MeanSrc} shows the relative rates of
mass supply from stellar winds and mass deposition due to local
thermal instabilities.  Notice that for $r < 1\,$kpc the mass sink is
switched off due to a stabilization by intensive SN-heating (see
equation \ref{GrowthRate}).  Mass dropout dominates on average at
radii from 2 to 3~kpc.  The oscillations of the mass dropout rate are
rather intense between 2 and 3.5~kpc, where the convective flow turns
over.

All of the best X-ray observations of elliptical galaxy X-ray halos
show positive temperature gradients in the central $\ge$10~kpc
\cite{brighenti.97}.  XMM-{\em Newton} observations of M87 and its
X-ray halo report $T\approx 1.3$~keV in the central circle of
$0.25^{\prime}$ ($\sim1.2$~kpc) in radius, a `steep' temperature rise
to $\sim2.3$~keV at $\sim3^{\prime}$ ($\sim15$~kpc), and a shallow
rise to $\sim2.9$~keV in the range from $3^{\prime}$ to $13^{\prime}$
\cite{boehringer+12.01}.  Yet the measured temperature gradients do
not give us much information about the temperature distribution within
the central kpc. Only the latter would help us to verify the model
assumptions concerning the thermal balance of the hot gas near the
centre.  The measured gradients can rather point to a slight
non-isothermality of the underlying gravitational potential, which
traces the transition from the dark matter halo/cluster dominated
outer region to the stellar core of M87 \cite{nulsen.95a,nulsen98}.

At the same time, the flat (on the average) entropy distribution in
the core region of our model is supported by the assumed spherical
symmetry of the underlying gravitational potential, average gas
density distribution, and the net heat source.  As we mention in the
introduction, this assumption is restrictive.  Even a small asymmetry,
e.g., in the central heating efficiency would create a hybrid
in/out-flow with a more stable (on the average) entropy
stratification, i.e. with a mild positive average temperature
gradient.  As soon as the asymmetric heating is sufficient and, still,
centrally concentrated (as in the case of AGN heating), such a flow
would not necessarily display a central density/X-ray brightness cusp.

Observationally, one can expect to see convection in the form of
uprising hot bubbles interacting with a background flow of cooling
gas.  It is possible, that flows of this kind are observed in X-rays
by {\em Chandra} and in radio on VLA in NGC~1275 \cite{fabian+8.00}
and in M87 \cite{owen..00}.  Such flows seem to be largely subsonic,
and the interface of a hot bubble with the background flow does not
imply shock waves unlike in the model of Heinz, Reynolds \& Begelman
\shortcite{heinz..98}.  Rather, it would be physically similar to the
interface of our convective core with the surrounding hydrostatic
corona, displaying lower temperatures, larger mass deposition rates
and a higher X-ray surface brightness.  We shall consider such flows
in detail elsewhere.

\subsubsection{Statistical properties: power spectra}\label{fourier}
\begin{figure*}
\vspace{-9.5cm}
\psfig{figure=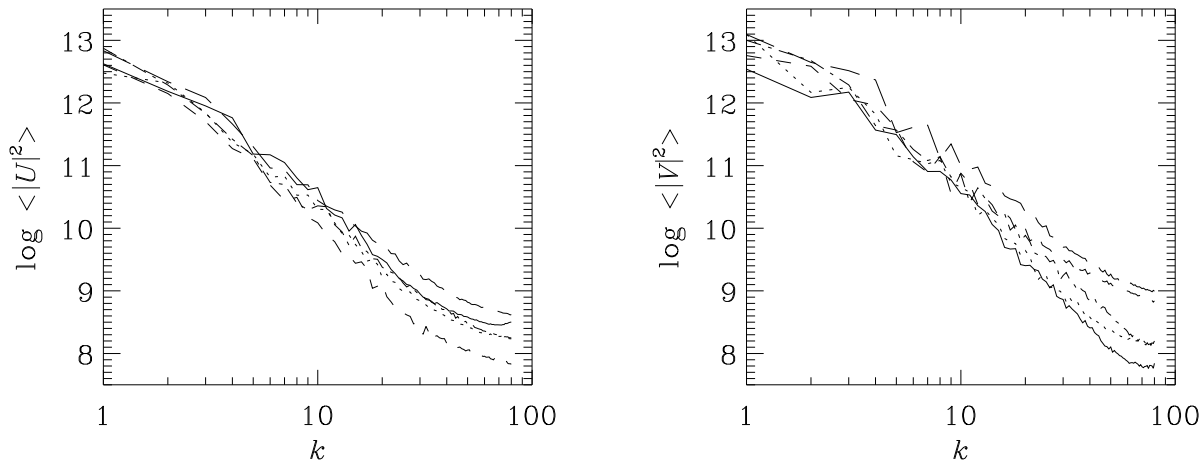,clip=}
\vspace{-9.cm}
\caption{Time averaged power spectra estimates of the vertical
($u$, left panel) and horizontal($v$, right panel) component of the
gas velocity.  The power spectra are shown for radii of 0.71 (solid
line), 1.32 (long dashes), 1.93 (dash dot), 2.54 (dashed), and
3.15~kpc (dotted), respectively.  The power spectra are averaged over
80 consecutive states covering the time interval 275--300~Myr.  The
spectra show no significant radial dependence.}
\label{Power}
\end{figure*}
A spectral decomposition of a flow helps to reveal information which
is not apparent otherwise, like the distribution of energy over length
scales and effects of the finite grid resolution. In order to reveal
this information we have computed one-dimensional Fourier transforms
along the $\theta$-coordinate direction at various radii.  Strong
gravitational stratification of the gas density and the lack of
periodicity in the radial direction would dominate the spectrum in the
radial direction, and thus would complicate the analysis.

For a scalar quantity of interest, $q$, we define the horizontal
Fourier transform to be
\begin{equation}
Q(r,k,t) = \frac{1}{\Delta\theta}\int^{\theta_R}_{\theta_L} 
q(r,\theta,t) e^{-2\pi ik\theta/\Delta\theta}d\theta,
\end{equation}
which is a function of time, radius, and the horizontal wavenumber
$k$.  Here $\Delta\theta$ is the horizontal periodic length.  The
power spectrum is the square of the modulus of its complex Fourier
transform, namely $\mid Q(r,k,t)\mid^2$.  We have used simple
periodograms [see, e.g., Press, Teukolsky, Vetterling \& Flannery
\shortcite{1992nrca.book.....P}] as the power spectra estimates for
discrete data sets given on the finest level grid covering the inner
convective region of the computational domain.

For the spectral analysis we have used the velocity field.  When the
density is constant, the power spectrum of the velocity field is
similar to a kinetic energy spectrum.  As the gas density varies only
slightly about its mean value with angle at a given radius (see $max$
and $min$ envelopes in Fig.~\ref{MeanFlow}), the density factor is
always negligible.  However, since radial modes are ignored in our 1D
analysis, the sum of the power spectra of the velocity components is
not expected to be a measure of the total kinetic energy.

Since instantaneous power spectra are noisy and since they fluctuate
considerably in time, we use temporal averages to estimate the typical
power distribution with wavenumber.  The convective core expands as
the source terms associated with the stellar system drive the hot
bubble at the centre.  Hence, we had to take averages over a period of
time shorter than the expansion time scale ($\sim10^8$~yr) but longer
than the lifetime of the largest mushroom-like structures in our
convective flow ($\sim20$~Myr, see Section \ref{integral}).  Averaging
over a time interval 25~Myr sufficiently suppresses the oscillations
in the mean spectra at small wavenumbers.  We use angle brackets
$<\ldots>$ in the following to denote the computed temporal averages.

Fig.~\ref{Power} shows the power spectra of the two velocity
components at radii of 0.71, 1.32, 1.93, 2.54, and 3.15~kpc,
respectively.  Obviously, the power is concentrated towards the large
scales, as already mentioned in section \ref{flowEvol} when discussing
the settling of the convective regime.  The spectra do not show any
significant radial dependence.  The mean slope $p$ of the spectra
$<\mid Q\mid^2>\propto k^p$ is $p \sim -\frac{5}{2}$.  Following the
work of Porter \& Woodward \shortcite{PW94}, who investigated the
numerical dissipation of the PPM scheme, we estimate that in our
simulations numerical dissipation starts to become important for
$k\ge10$, i.e. for structures smaller than about 16 angular zones.

Batchelor (1969) \nocite{batchelor69} in his paper on two-dimensional
isotropic turbulence derived an analogue of the Kolmogorov energy
spectrum. For 2D flows the spectrum of vorticity  
\mbox{$\frac{1}{2}<(\nabla\times v)^2>$} 
is universal. If $\Omega(k)$ is the vorticity
spectrum, then $\Omega(k)\propto k^{-1}$ is the analogue of
Kolmogorov's power law. For the energy spectrum of 2D flows one has
$E(k) \propto k^{-2} \Omega(k) \propto k^{-3}$.  From our Fourier
analysis we obtain as expected $|U^2| \propto k^{-3}$ and
$|V^2|\propto k^{-3}$ in the interval $k\in[3, 30]$. This explains the
transition from small scale to large scale mushrooms when convection
sets in.

\subsubsection{Average transport properties}
The characteristics of the convective flow in the inner region can
also be examined by calculating the angular averaged energy and a set
of energy fluxes, which we define below following Hurlburt, Toomre \&
Massaguer \shortcite{1986ApJ...311..563H}.

The angular integrated total energy equation (\ref{Energy}) can be
expressed as
\begin{equation}
\frac{\partial \hat{E}}{\partial t} + 
\frac{1}{r^2}\frac{\partial }{\partial r}r^2(
F_{\rmn{E}} + F_{\rmn{C}} + F_{\rmn{K}}) = S,  
\end{equation}
with
\begin{equation}
\hat{E} = 2\pi\int_{\theta_L}^{\theta_R} E\:r^2\sin{\theta}\:d\theta,
\end{equation}
where $F_{\rmn{E}}$ is the heat flux due to a spherically symmetric
expansion or contraction of the model, $F_{\rmn{C}}$ is the `enthalpy'
or convective heat flux, $F_{\rmn{K}}$ is the kinetic heat flux, and
$S$ is the net energy source term.  These fluxes and the source term
are defined as
\begin{equation}
F_{\rmn{E}} = 
2\pi\int_{\theta_L}^{\theta_R}
\overline{u \rho}\cdot\left(
\overline{e + \frac{p}{\rho}+\frac{\bmath{v}^2}{2}} - 
\frac{d\phi}{dr}\right) \:r^2\sin{\theta}\:d\theta,
\end{equation}
\begin{equation}
F_{\rmn{C}} =
2\pi\int_{\theta_L}^{\theta_R}
u \rho\cdot\left(e + \frac{p}{\rho}\right)^{\prime}
\:r^2\sin{\theta}\:d\theta,
\end{equation}
\begin{equation}
F_{\rmn{K}} = 
2\pi\int_{\theta_L}^{\theta_R}
u \rho\cdot\left(\frac{\bmath{ v}^2}{2}\right)^{\prime}
\:r^2\sin{\theta}\:d\theta,
\end{equation}
\begin{equation}
S =
2\pi\int_{\theta_L}^{\theta_R}
\-\left[
\alpha\rho_* e_* - \dot{\rho}_{ti}\left(\gamma e + \frac{\bmath{v}^2}{2}
\right) - \rho^2\Lambda
\right]
\:r^2\sin{\theta}\:d\theta.
\end{equation}
The above definitions involve spatially fluctuating variables $f =
\overline{f} + f^{\prime}$, where $\overline{f}$ is the angular mean
value and $f^{\prime}$ is the fluctuation about that mean value. They
allow us to discuss correlations between the radial velocity and a
number of fluctuating quantities.  Such correlations provide useful
insights into energy generation and energy transport mechanisms in the
convective interior of the hot ISM.

We further define a total energy flux $F_{\rmn{T}} = F_{\rmn{E}} +
F_{\rmn{C}} + F_{\rmn{K}}$.  In a steady state with $S=0$ the total
flux does not depend on radius $r$.  If a convective flow achieves
a statistically steady state, then for sufficiently long time averages
the average total energy flux
\begin{equation}
<F_{\rmn{T}}>\; =\; <F_{\rmn{E}}>+<F_{\rmn{C}}>+<F_{\rmn{K}}> 
\end{equation}
does not dependent on radius as well, while the individual
contributing fluxes may vary considerably with $r$.
\begin{figure}
\vspace{-4.7cm}
\psfig{figure=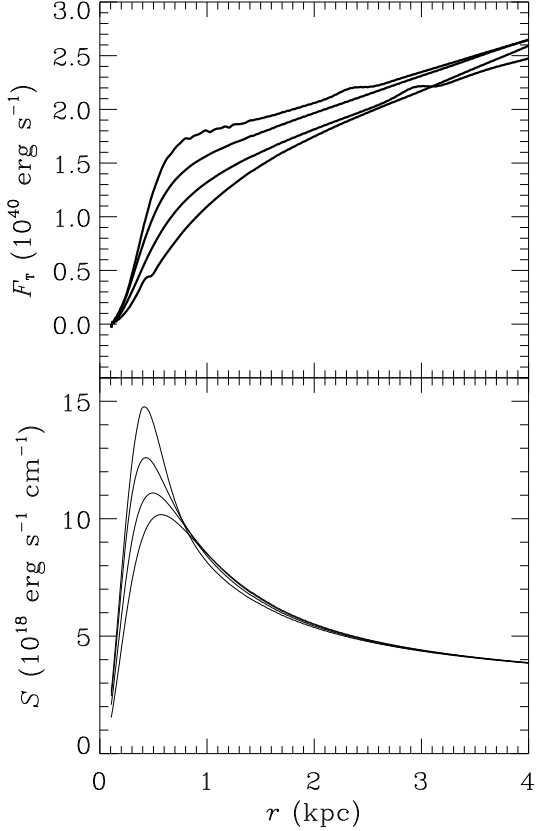,clip=}
\vspace{-9.cm}
\caption{Total angular integrated energy flux $F_{\rmn{T}}$ (upper
panel) and net energy source $S$ (lower panel) at $t= 10$, 30, 50, and
70~Myr (lines from top to bottom in each panel), respectively.}
\label{InitialFluxes}
\end{figure}
Before turning to the fluctuations let us consider the inner 4\,kpc
region where convective motion occurs in a mean `background' flow
after the onset of the instability.  Fig.\,\ref{InitialFluxes} shows
the radial dependence of the total energy flux $F_{\rmn{T}}$ and net
source $S$ at four representative moments of time before and shortly
after the instability sets in.  At these times the total flux is
directed outwards and increases from zero at $r=0$ to values around
$2.5 \times 10^{40}\,$erg/s at $r =4\,$kpc mainly due to a slow
overall expansion of the gas.  This expansion is stimulated partly by
an outgoing weak sound wave generated by the initially imperfect
hydrostatic equilibrium, and also by the increasing source terms
blowing a hot bubble at the centre.  Since the flow is spherically
symmetric for $t \le 50$~Myr, the major contribution to the total flux
is due to gas expansion in an external gravity field,
i.e. $F_{\rmn{T}} = F_{\rmn{E}}$.  The small wrinkles visible at $r
\approx 1\,$kpc in the flux diagram at $t = 70\,$Myr hint towards a
contribution from fluctuations of the developing convective flow, but
still $F_{\rmn{T}}\approx F_{\rmn{E}}$.
\begin{figure}
\vspace{-4.7cm}
\psfig{figure=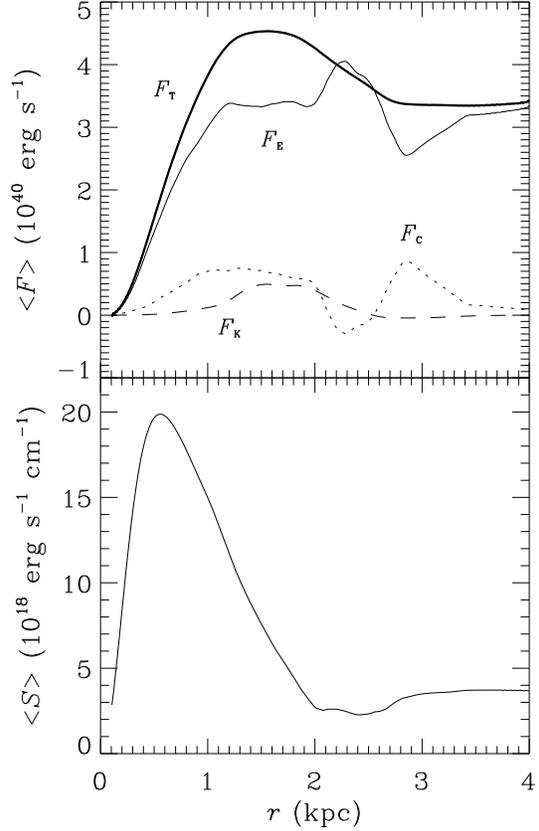,clip=}
\vspace{-9.cm}
\caption{Mean energy fluxes $<F>$ (upper panel) and mean net energy
source $<S>$ (lower panel) averaged over $295.5 < t < 308.3$~Myr.  The
heavy solid line shows the mean total flux $<F_{\rmn{T}}>$, and the
solid, dotted, and dashed lines show the heat flux due to a
spherically symmetric expansion or contraction of the model
$<F_{\rmn{E}}>$, the convective flux $<F_{\rmn{C}}>$, and the kinetic
flux $<F_{\rmn{k}}>$, respectively.}
\label{Fluxes}
\end{figure}
By $t=300$~Myr, when the convective flow already covers the whole
inner region ($r\leq3$~kpc), the global expansion driven by the action
of the source terms remains the dominant contributor to the total
energy flux (Fig.~\ref{Fluxes}).  Note that the mean energy flux
carried away from the inner region at $t \sim 300$~Myr amounts to
about 7 per cent of the luminosity of the hot gas in the whole
computational domain $r\le400$~kpc (see Fig.~\ref{Integral}).

Our analysis indicates that the fluctuations play only a secondary
role for the energy transport.  The mean angular integrated
convective flux, $<F_{\rmn{C}}>$, is positive everywhere except for
the region around $r \approx 2.3$~kpc, where the mean source $<S>$
attains its minimum, yielding an upward convective transport of
thermal energy.  Like in case of an incompressible fluid a positive
flux value corresponds to motions where hot fluid parcels rise and
cold ones sink.  The depression in $<F_{\rmn{C}}>$ at $r \approx
2.3$~kpc spans the same range in radius as the condensation front
smeared by convection.  In this region mass dropout due to thermal
instabilities dominates over mass supply from stellar winds, and the
enthalpy flux, carried by fluctuations from the inner region,
decreases.

The mean kinetic flux $<F_{\rmn{K}}>$ steadily grows in absolute
value, as the convective core becomes more extended giving rise to
larger convective structures with higher velocities.  The kinetic flux
is positive (i.e. directed outwards) at $r \le 2.5$~kpc and weakly
negative further out in the region where convective motions overshoot
through the condensation front.  The sign of $F_{\rmn{K}}$ is related
to differences in strength and filling factor between upward and
downward directed motions.  In our two-dimensional simulations upflows
are considerably faster than downflows in the interior of the
convective core and there is no apparent asymmetry in the filling
factors (see the angular velocity diagram in Fig.~\ref{MeanFlow}).

\subsubsection{Integral characteristics\label{integral}}
\begin{figure}
\vspace{-4.7cm}
\psfig{figure=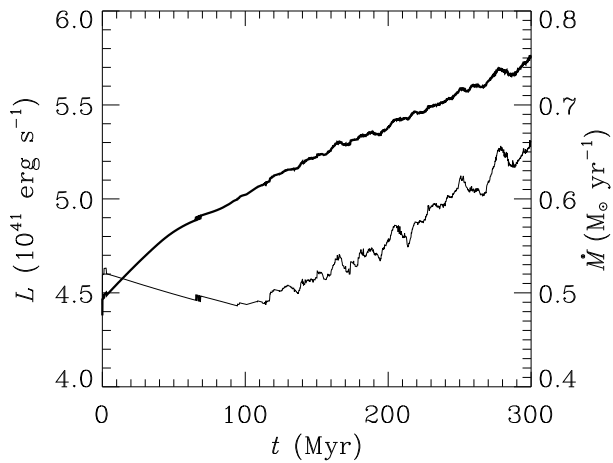,clip=}
\vspace{-14.cm}
\caption{Temporal behaviour of the luminosity $L$ (thin curve) and net
mass deposition rate $\dot M$ (thick curve) integrated over the whole
computational domain.}
\label{Integral}
\end{figure}
In order to follow the evolution of global characteristics of the flow
we computed the total hot gas luminosity $L$ and the net mass
deposition rate $\dot M$ as integrals over the whole computational
domain (Fig.\,\ref{Integral}).  The computed total luminosity is close
to the X-ray luminosity $L_{\rmn{X}} \approx 7.2\times10^{41}$~erg
s$^{-1}$ of NGC~4472 at a distance of 17~Mpc.  There is a mild
tendency for the luminosity to grow with time as the gas shed by stars
in the galaxy slowly accumulates in the convective core.  Low
amplitude oscillations in the luminosity on a time scale of $\tau \sim
20-30\,$Myr correspond to a life cycle of the largest convective
structures.  Assuming a linear scale of $l \sim 3\,$kpc and $\tau =
25\,$Myr, one can estimate a characteristic velocity $l / \tau \approx
120\,$km\,s$^{-1}$.  This value is typical for the convective core
which developed in the simulations by $t \sim 300\,$Myr (see
Fig.~\ref{MeanFlow}).  A time scale of 25\,Myr is also a reasonable
choice for the time averaging interval to be used when computing mean
power spectra.

The net mass deposition rate $\dot M$, given by the volume integral 
\begin{equation}
\dot M = \int_V(\alpha\rho_* - \dot\rho_{\rmn{ti}})dV\,,
\end{equation}
is initially positive, since a negative global density perturbation
$\delta \rho/\rho = -0.1$ breaks the balance between the mass supply
to the hot phase from stellar winds and mass loss from the hot phase
due to local thermal instabilities.  The mass supply from stars always
dominates, and $\dot M$ grows from 0.5 to 0.75\,M$_{\odot}$\,yr$^{-1}$
(Fig.\,\ref{Integral}) since the system evolves further away from
hydrostatic equilibrium.  For our model galaxy with a stellar mass
$M_* \approx 8.5 \times 10^{11}\,$M$_{\odot}$ the rate of steady
supply of gas from stars, $\alpha_* M_* \approx
1.2\,$M$_{\odot}$\,yr$^{-1}$, implies that the gas condensation rate
varies from 0.7 at $t=0$ to $0.45\,$M$_{\odot}$ yr$^{-1}$ at 300\,Myr.
As one can see in Fig.~\ref{Integral}, $\dot M$ slightly oscillates in
phase with the X-ray luminosity.  Maxima of $\dot M$ record moments
when streams of rising hot material form new large scale mushroom-like
structures.

\section{Discussion and Conclusions}
In this paper we have studied hot gas flows originating in the central
few kpc of hot galactic coronae in response to a slight energy supply
imbalance in a model initially in equilibrium.  It is shown that a
negative energy budget drives a quasi-steady compact cooling inflow,
which is stable to small perturbations.  Excessive heating creates an
isentropic (on average) convective core in the vicinity of the galaxy
centre.  Provided the net energy gain in the core is not too high, the
convection remains subsonic with typical flow velocities (both, radial
and angular) of the order of $\pm 150\,$km\,s$^{-1}$.  The
characteristic convective pattern consists of a manifold of
mushroom-like structures of different scales.  The power spectra of
the velocity components are dominated by the largest-scale mushrooms.
The emission weighted gas temperature in the core is slightly rising
towards the centre.  Unlike in the case of a cooling inflow, the X-ray
surface brightness of a convective core does not display a sharp
maximum at the centre.  The inner parts of the convective core are
dominated by mass supply from evolved stars, while in the colder outer
shell mass drop out dominates the mass budget.  Large-scale mushrooms
(mainly their caps) can probably be marginally detected by high
resolution X-ray images with good photon statistics, as their volume
emissivity is about a factor of 2 higher than that of the surrounding
gas.

We have considered an idealized case with a spherically symmetric heat
source and small deviations from `thermal' equilibrium near the
centre.  However, we can extrapolate our results to a more realistic
hydrodynamic picture.  First, stronger heating in the vicinity of the
centre can drive a supersonic convection with velocities in excess of
400\,km\,s$^{-1}$ and produce a much larger (here 3-4\,kpc) convective
region.  Second, a Type Ia SN heating rate which is lower all over the
galaxy, and an excessive heating by an AGN in the central region will
create a hybrid flow with an inner convective core and an outer
`cooling flow', which will vanish in convection when the inflow
velocity is very subsonic.  Our preliminary results indicate that such
a hybrid dynamic flow can be stable and can last as long as the central
source is active.  Third, a low Type Ia SN heating rate without extra
energy supply near the centre will drive a more extended massive
cooling inflow with an even more pronounced central surface brightness
peak.  It would be harder to turn such a dynamic inflow into an
outflow by a strong perturbation (e.g., a burst of AGN activity, or a
merger) than a weaker compact inflow.  Finally, we expect that an
asymmetric central heat source in excess of the equilibrium value
would create bubble(s) filled with a (mean) isentropic convective flow
with properties similar to the ones described above.

{\em Chandra}, {\em Newton} and future X-ray missions will teach us
more about the hydrodynamics of `cooling flows'.

\section*{Acknowledgments}
AK was partly supported by the Russian Foundation for Basic Research
(project 98-02-19670) and by the Federal Targeted Programme {\em
Integration} (project A~0145).  TP was partly supported by grant
2P03D.014.19 from the Polish Committee for Scientific Research. The
simulations were performed on the RISC cluster of the MPA, and on the
SGI Power Challenge at the Interdisciplinary Centre for Computational
Modelling in Warsaw.

\bsp
\label{lastpage}


\begin{thebibliography}{}
%
%
%
\bibitem[Arimoto et al.\ 1997]{1997ApJ...477..128A} 
Arimoto N., 
Matsushita K., Ishimaru Y., Ohashi T. \& Renzini A., 1997, \apj, 477, 
128 

\bibitem[\protect\citename{Batchelor }1969]{batchelor69}
        Batchelor G. K., 1969,
        Phys. Fluids Suppl. II, 233


\bibitem[\protect\citename{Berger \& Colella }1989]{BC89}
        Berger M. J., Colella P.,
        1989,
        J. Comput. Phys., 82, 64

\bibitem[\protect\citename{Binney \& Tabor }1995]{binney.95}
        Binney J., Tabor G.,
        1995,
 \mnras, 276, 663

\bibitem[\protect\citename{B{\"o}hringer et al. }2001]{boehringer+12.01}
	B{\"o}hringer H., Belsole E., Kennea J., Matsushita K., 
	Molendi S., Worrall D. M., Mushotzky R. F.,
        Ehle M., Guainazzi M., Sakelliou I., Stewart G.,
        Vestrand W. T., Dos Santos S., 2001,
 A\&A, 365, L187

\bibitem[\protect\citename{Brighenti \& Mathews }1997]{brighenti.97}
        Brighenti F., Mathews W. G.,
        1997,
 \apj, 486, L83

\bibitem[\protect\citename{Buote }1999]{1999MNRAS.309..685B} 
Buote D.\ A., 1999, \mnras, 
309, 685 


\bibitem[\protect\citename{Cappellaro { et~al.} }1999]{cappellaro..99}
Cappellaro E., Evans R., Turatto M., 1999,
\newblock A\&A, {351}, 459


\bibitem[\protect\citename{Colella \& Woodward }1984]{CW84}
        Colella P., Woodward P. R., 1984,
        J. Comput. Phys., 59, 264

\bibitem[\protect\citename{Davis \& White }1996]{davis.96}
Davis D.S., White R.E., III, 1996,
\newblock ApJ, {470}, L35

\bibitem[\protect\citename{Faber \& Gallagher }1976]{faber.76b}
Faber S.~M., Gallagher J.~S., 1976,
\newblock ApJ, {204}, 365

\bibitem[\protect\citename{Fabian et al. }2000]{fabian+8.00}
Fabian A. C., Sanders J. S., Ettori S., Taylor G. B., Allen S. W., 
Crawford C. S., Iwasawa K., Johnstone R. M., Ogle P. M., 2000,
 \mnras, 318, L65

\bibitem[Fujita, Fukumoto \& Okoshi 1997]{1997ApJ...488..585F} 
Fujita Y., Fukumoto J. \& Okoshi K., 1997, \apj, 488, 585 

\bibitem[Harris et al. 1999]{harris...1999} 
Harris D.E., Owen F.N., Biretta J.A. \& Junor W., 1999,
in H.~B\"ohringer, L.~Feretti \& P.~Shuecker (eds), 
Diffuse Thermal and Relativistic Plasma, 
MPE Report 271, 111 

\bibitem[\protect\citename{Heinz, Reynolds \& Begelman }1998]{heinz..98} 
Heinz S., Reynolds C. S., Begelman M. C., 1998, \apj, 501, 126 

\bibitem[\protect\citename{Hunter }1970]{hunter70} 
Hunter J. H., Jr., 1970,  \apj, 161, 451

\bibitem[\protect\citename{Hurlburt, Toomre \& Massaguer }1986]
{1986ApJ...311..563H}  
Hurlburt N.\ E., Toomre J.\ \& Massaguer J.\ M., 1986, \apj, 311, 563 

\bibitem[\protect\citename{Inogamov }1999]{inogamov99}
Inogamov N.A., 1999,
\newblock Astrophys. \& Space Phys. Rev., {10-2}, 1

\bibitem[\protect\citename{Kritsuk }1992]{kritsuk92}
Kritsuk A.~G., 1992,
\newblock A\&A, {261}, 78

\bibitem[\protect\citename{Kritsuk }1996]{kritsuk96}
Kritsuk A.~G., 1996,
\newblock MNRAS, {280}, 319

\bibitem[\protect\citename{Kritsuk }1997]{kritsuk97}
Kritsuk A.~G., 1997,
\newblock MNRAS, {284}, 327

\bibitem[\protect\citename{Kritsuk et~al. }1998]{kritsuk..98}
	Kritsuk A.~G., B\"ohringer H., M\"uller E., 1998,
	\newblock MNRAS, {301}, 343

\bibitem[Koekemoer et al.\ 1999]{1999ApJ...525..621K} 
Koekemoer A.\ M., 
O'Dea C.\ P., Sarazin C.\ L., McNamara B.\ R., Donahue M., Voit G.\ 
M., Baum S.\ A. \& Gallimore J.\ F., 1999, \apj, 525, 621 

\bibitem[\protect\citename{Loewenstein }1997]{1997gccf.conf..100L} 
Loewenstein M., 1997, 
in N.~Soker (ed.), Galactic and Cluster Cooling Flows, 
ASP Conf.\ Ser.\ 115, 100 

\bibitem[\protect\citename{Loewenstein \& Mushotzky } 1998]{1998IAUS..188...53L} 
Loewenstein M. \& Mushotzky R.\ F., 1998, IAU Symp.\ 188: The Hot 
Universe, 188, 53 

\bibitem[McNamara et al.\ 2000]{2000ApJ...534L.135M} 
McNamara B.\ R.\ et al.,  2000, \apj, 534, L135 

\bibitem[\protect\citename{Mathews }1990]{mathews90}
Mathews W.G., 1990,
\newblock ApJ, 354, 468


\bibitem[\protect\citename{Mathews \& Brighenti }1999]{mathews.99}
Mathews W.G., Brighenti,~F., 1999,
\newblock ApJ, 526, 114

\bibitem[\protect\citename{Nulsen }1998]{nulsen98}
Nulsen P.E.J.,
\newblock {MNRAS}, 1998, 297, 1109

\bibitem[\protect\citename{Nulsen \& {B\"ohringer} }1995]{nulsen.95a}
Nulsen P.E.J., {B\"ohringer} H.,
\newblock {MNRAS}, 1995, 274, 1093


\bibitem[\protect\citename{Owen et al. }2000]{owen..00}
Owen F. N., Eilek J. A., Kassim N. E., 2000,
\newblock ApJ, 543, 611

\bibitem[\protect\citename{Plewa }1993]{P93}
        Plewa T.,
        1993,
        Acta Astr., 43, 235


\bibitem[\protect\citename{Plewa \& M\"uller }2000]{PM00}
        Plewa T., M\"uller E., 2001,
        Comp. Phys. Comm., in press


\bibitem[\protect\citename{Porter \& Woodward }1994]{PW94}
        Porter D.H, Woodward P.R., 1994,
        ApJS, 93, 309

\bibitem[\protect\citename{Press, Teukolsky, Vetterling \& Flannery }1992]
{1992nrca.book.....P} 
	Press W.\ H., Teukolsky S.\ A., Vetterling 
	W.\ T.\ \& Flannery B.\ P., 1992, Numerical recipes in C. The
	art of scientific computing, 
	Cambridge: University Press 

\bibitem[\protect\citename{Sarazin \& White }1987]{sarazin.87}
Sarazin C.~L., White R.~E., III, 1987,
\newblock ApJ, {320}, 32

\bibitem[\protect\citename{Sutherland \& Dopita }1993]{SD93}
        Sutherland R. S., Dopita M. A., 1993,
        ApJS, 88, 253

\bibitem[\protect\citename{Thomas }1986]{thomas86}
Thomas P.~A., 1986,
\newblock MNRAS, {220}, 949

\bibitem[\protect\citename{Thomas }1988]{1988cfcg.work..235T} 
Thomas, P.\ A.\ 1988, in A.C.Fabian (ed.), Cooling 
Flows in Clusters and Galaxies, 235 

\bibitem[\protect\citename{van~den Bergh \& Tammann }1991]{vandenbergh.91}
van~den Bergh S., Tammann G.~A., 1991,
\newblock ARA\&A, {29}, 363

\end{thebibliography}
\end{document}